\documentclass[%
 aip,
 amsmath,amssymb,
 reprint,%
]{revtex4-1}

\usepackage{graphicx}
\usepackage{dcolumn}
\usepackage{bm}
\usepackage{xcolor}
\usepackage[utf8]{inputenc}
\usepackage[T1]{fontenc}
\usepackage{mathptmx}
\usepackage{etoolbox}
\usepackage[normalem]{ulem}
\makeatletter
\def\@email#1#2{%
 \endgroup
 \patchcmd{\titleblock@produce}
  {\frontmatter@RRAPformat}
  {\frontmatter@RRAPformat{\produce@RRAP{*#1\href{mailto:#2}{#2}}}\frontmatter@RRAPformat}
  {}{}
}%
\makeatother
\begin{document}

\preprint{AIP/123-QED}

\title{High detectivity terahertz radiation
sensing using frequency-noise-optimized nanomechanical resonators 
}

\author{Chang Zhang}
 \affiliation{Department of Mechanical Engineering, University of Ottawa, Ottawa, ON, Canada}
\author{Eeswar K. Yalavarthi}
 \affiliation{Department of Physics, University of Ottawa, Ottawa, ON, Canada}
 
\author{Mathieu Giroux}
 \affiliation{Department of Mechanical Engineering, University of Ottawa, Ottawa, ON, Canada}

\author{Wei Cui}
 \affiliation{Department of Physics, University of Ottawa, Ottawa, ON, Canada}
\author{Michel Stephan}
 \affiliation{Department of Mechanical Engineering, University of Ottawa, Ottawa, ON, Canada}

\author{Ali Maleki}
 \affiliation{Department of Physics, University of Ottawa, Ottawa, ON, Canada}

\author{Arnaud Weck}
 \affiliation{Department of Mechanical Engineering, University of Ottawa, Ottawa, ON, Canada}

\author{Jean-Michel Ménard}
 \affiliation{Department of Physics, University of Ottawa, Ottawa, ON, Canada}

\author{Raphael St-Gelais}
 \affiliation{Department of Mechanical Engineering, University of Ottawa, Ottawa, ON, Canada}
 \email{raphael.stgelais@uottawa.ca}

\date{\today}

\begin{abstract}
We achieve high detectivity terahertz radiation sensing using a silicon nitride nanomechanical resonator functionalized with a metasurface absorber. High performances are achieved by striking a balance between the frequency stability of the resonator, and its responsivity to absorbed radiation. Using this approach, we demonstrate a detectivity $D^*\approx3.4\times10^9~\mathrm{cm\cdot\sqrt{Hz}/W}$ and a noise equivalent power $\mathrm{NEP}\approx36~\mathrm{pW/\sqrt{Hz}}$ that outperform the best room-temperature on-chip THz detectors, such as pyroelectric detectors, while maintaining a comparable thermal response time of $\approx200$ ms. Our optical absorber consists of a 1-mm diameter metasurface, which currently enables a 0.5 -- 3 THz detection range but can easily be scaled to other frequencies in the THz and infrared ranges. In addition to demonstrating high-performance terahertz radiation sensing, our work unveils an important fundamental trade-off between frequency stability and responsivity in thermal-based nanomechanical radiation sensors.

\end{abstract}

\maketitle
\section{Introduction}
Although thermal-based sensors have been used for decades for incoherent long-wavelength (infrared and terahertz) detection, their performance still falls significantly short of the fundamental detectivity limit imposed by the photon fluctuation noise of a blackbody\cite{ref1} ($D^*=1.8\times10^{10}~\mathrm{cm\cdot\sqrt{Hz}/W}$). This performance gap largely results from non-fundamental imperfections, such as Johnson-Nyquist electrical noise, occurring in existing thermal-based sensing elements. In recent years, temperature-sensitive nanomechanical resonators have been proposed as a promising alternative. Measuring temperature changes with a mechanical resonance can in-principle be made immune from electrical noise, using for example an optical-based de-coupled readout. Resonators made of thin-film materials e.g., silicon nitride (SiN)\cite{ref2,ref3,ref4,ref5,ref6}, aluminum nitride (AlN) \cite{ref7}, graphene \cite{ref8}, gallium arsenide (GaAs) \cite{ref9} have been investigated extensively for thermal-based radiation sensing at infrared wavelengths. Using a similar thermal-based sensing approach, resonators coated with additional metal absorbers have been proposed for detection at THz \cite{ref10} (0.25–3 THz) and sub-THz \cite{ref11} (0.1–0.3 THz) frequencies. 

A common approach for optimizing performances in these nanomechanical sensors is by maximizing the magnitude of the mechanical resonance frequency shift relative to optical power absorption (i.e., maximizing thermal responsivity $R$). This is typically done by utilizing resonators of very small effective mass (i.e., effective length of the sensing area from $10^1~\mathrm{\mu m}$ to $10^2~\mathrm{\mu m}$ \cite{ref2,ref3,ref4,ref8}) and by thermally isolating them via extremely thin supporting structures (e.g., tether \cite{ref6,ref8,ref11}, rod \cite{ref2,ref4}, etc). Surprisingly, resonators frequency stability is typically not as central to the design process as responsivity $R$, even though it is equally important to the determination of the final noise figure. Considering recent studies on frequency noise in nanomechanical resonators \cite{ref12,ref13,ref14,ref15,ref16}, we find that approaches for improving the responsivity $R$, such as mass miniaturization and thermal isolation enhancement, can also significantly degrade resonators frequency stability and hence the overall sensing performance. As a result, the specific detectivity $D^*$ of recently reported resonator-based THz \cite{ref10} and sub-THz \cite{ref11} detectors still falls short of the best commercial room-temperature, on-chip THz detectors (i.e., pyroelectric detectors \cite{ref17,ref18} with $D^*\approx7\times10^8~\mathrm{cm\cdot\sqrt{Hz}/W}$) by factors 68 and 2, respectively. Likewise, their performances are at least one order of magnitude below those of Golay cells \cite{ref19} ($D^*\approx4\times10^9~\mathrm{cm\cdot\sqrt{Hz}/W}$).

Here, we show that maximizing responsivity by miniaturization can be a counterproductive sensor design approach, and that greater detection performance gains can be realized by improving frequency stability using a resonator of relatively large mass. By striking a balance between responsivity and frequency stability, we experimentally demonstrate an uncooled THz detector with $\mathrm{NEP}\approx36~\mathrm{pW/\sqrt{Hz}}$ and specific detectivity $D^*\approx3.4\times10^9~\mathrm{cm\cdot\sqrt{Hz}/W}$ at 2 THz incident radiation frequency. Our approach therefore exhibits two orders of magnitude improvement in $D^*$ compared with resonator-based detectors operating in THz frequency range \cite{ref10} and a factor of 5 improvement in $D^*$ compared with the highest performance commercial room-temperature, on-chip THz detectors \cite{ref17,ref18}.

\section{Intrinsic noise in thermal-based nanomechanical radiation sensing}

Before diving into the specifics of our nanomechanical THz detector, we outline important parameters of a square SiN membrane resonator for thermal-based detection of radiation (i.e., at any incident wavelength). For simplicity, this theoretical investigation assumes uniform absorption of incident radiation across the entire membrane area. Adjustments for a localized absorber are discussed in Section IV. We first define noise equivalent power (in $\mathrm{W/\sqrt{Hz}}$) as:
\begin{equation}
\mathrm{NEP}=\frac{\sqrt{S_y}}{R},
\label{eq:1}
\end{equation}
where $R$ is the sensor responsivity of the fractional frequency shift to incident radiation (in $\mathrm{W^{-1}}$), $S_y$ is the resonator fractional frequency noise spectral density (in $\mathrm{Hz^{-1}}$) for a given eigenmode of frequency $f_r$. We express all noise spectral densities and filters in $\mathrm{Hz}$ rather than in $\mathrm{rad/s}$, as detailed in Supplementary Section S1. This allows for better uniformity with noises in conventional electronic instrumentation (e.g., photodetectors and bolometers) which are more commonly expressed in this unit base.

The frequency noises $S_y$ of a typical nanomechanical resonator can be categorized into intrinsic noises and extrinsic detection noise (i.e., due to nonideality in readout methods). For this theoretical analysis, we investigate a temperature-sensitive, high-Q factor resonator without detection noise. The total noise profile $(S_y = S_{y,TF}+S_{y,TM})$ therefore comprises thermomechanical $S_{y,TM}$ and thermal fluctuation $S_{y,TF}$ noises (i.e., with intrinsic noises only). Thermomechanical noise (in $\mathrm{Hz^{-1}}$) is given by \cite{ref21}:
\begin{equation}
S_{y,TM}(f) = \frac{k_BT}{4\pi^2m_{eff}f_r^3QA_{rss}^2}\Big|H_{mech}(f)\Big|^2,
\label{eq:2}
\end{equation}
where $k_B$ is the Boltzmann constant, $T$ is the background environment temperature, $m_{eff}$ is the resonator effective mass, $A_{rss}$ is the driven oscillation amplitude, $H_{mech}(f)=1/(1+j2\pi f \tau_{mech})$ is a one-pole low-pass filter that accounts for the intrinsic mechanical response of the resonator, where $\tau_{mech}=Q/\pi f_r$ is the mechanical time constant of the resonator, and $Q$ is the mechanical quality factor at eigenfrequency $f_r$. In turn, thermal fluctuation noise (in $\mathrm{Hz^{-1}}$) is given by\cite{ref16,ref22,ref23}:
\begin{equation}
S_{y,TF}(f) = \frac{4k_BT^2\alpha^2}{G}\Big|H_{th}(f)\Big|^2,
\label{eq:3}
\end{equation}
where $G$ (in $\mathrm{W/K}$) is the total thermal conductance between the resonator and its environment and $H_{th}(f)=1/(1+j2\pi f \tau_{th})$ is a one-pole low-pass filter that accounts for the intrinsic thermal response of the resonator, where $\tau_{th}$ is the thermal time constant. $\alpha$ is the temperature coefficient of fractional frequency shifts (in $\mathrm{K^{-1}}$). 

The responsivity of the fractional frequency shift to incident radiation (in $\mathrm{W^{-1}}$) can be expressed as:
\begin{equation}
R = \frac{\gamma \alpha}{G_{eff}}\Big|H_{th,eff}(f)\Big|,
\label{eq:4}
\end{equation}
where $\gamma$ is the absorption coefficient at the target detection wavelength (e.g., $\gamma=0.4$ at $\mathrm{2~ THz}$ in this work). $G_{eff}$ and $H_{th,eff}$ are the effective thermal conductance and thermal filter. They are equal to $G$ and $H_{th}$ for uniform absorption of radiation across the entire membrane area. We assume $G_{eff}=G$ in this section, and re-evaluate them using finite element analysis for the membrane functionalizes with a localized absorber in Section IV.

In the limit case where thermomechanical noise is minimized relative to thermal fluctuation noise, we can substitute  Eq.~(\ref{eq:3}) and (\ref{eq:4}) into Eq.~(\ref{eq:1}) which yields the performance limit:

\begin{align}
\mathrm{NEP_{TF}}& =\frac{\sqrt{S_{y,TF}}}{R} = \frac{\sqrt{4k_BT^2G}}{\gamma}.
\label{eq:5}
\end{align}
It is worth noting that in this limit, $\mathrm{NEP}$ becomes independent of both the intrinsic thermal filter $H_{th}$ and the temperature coefficient $\alpha$.

Since many noise sources scale with detector area $A_{det}$, the smallest detectable radiation intensity (in $\mathrm{W/m^2}$) is often a more universal metric than the smallest measurable power (in $\mathrm{W}$). Specific detectivity $D^*=\sqrt{A_{det}}/\mathrm{NEP}$ is therefore useful for quantifying the ultimate performance limit of detectors. The limit case for $D^*$ occurs when $S_y = S_{y,TF}$ and all heat transfer occurs via radiation, yielding
\begin{equation}
G=G_{rad}=4A_{rad}\sigma_{SB}\varepsilon T^3,
\label{eq:7}
\end{equation}
where $A_{rad}=2L^2$ accounts for radiative heat transfer on both sides of the membrane. By substituting $G_{rad}$ into Eq. (\ref{eq:5}), we obtain the expression of $D^*$ limited solely by thermal photon fluctuation:
\begin{align}
D^*_\mathrm{TF,~photon}& =\frac{\sqrt{A_{det}}}{\mathrm{NEP_{TF}}|_{G=G_{rad}}} = \frac{\gamma}{\sqrt{32k_B \sigma_{SB}T^5\varepsilon}}.
\label{eq:8}
\end{align}

For the case of a blackbody thermal detector (i.e., $\gamma=\varepsilon=1$), and for $A_{det}=L^2$, Eq.~(\ref{eq:8}) yields $D^*_\mathrm{TF,~photon}\approx1.3\times10^{10}~\mathrm{cm\cdot\sqrt{Hz}/W}$. This is slightly different from the commonly known fundamental limit $D^*\approx1.8\times10^{10}~\mathrm{cm\cdot\sqrt{Hz}/W}$ for thermal detectors \cite{ref1}, that assumes thermal radiation occurring on one side of the detector only (i.e., $A_{det}=A_{rad}$). Furthermore, considering our specific case of a $\mathrm{THz}$ absorbing membrane resonator in which $\gamma=0.4$ in the $\mathrm{THz}$ range and $\varepsilon=0.1$ in the thermal range, the limit for the sensor in this work is  $D^*_\mathrm{TF,~photon}\approx1.6\times10^{10}~\mathrm{cm\cdot\sqrt{Hz}/W}$.

\begin{figure*}
\includegraphics[scale=1.13]{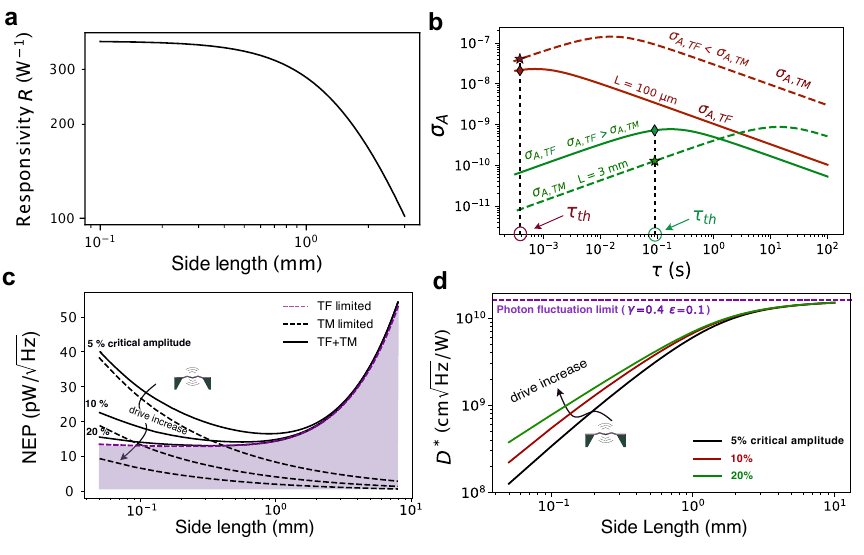}
\caption{\label{fig:1}(a) Thermal responsivity of fractional frequency shifts for square SiN membranes of various lengths $L$, and a fixed thickness of 90 nm. (b) Theoretical Allan deviations $\sigma_A$ of two SiN membranes with significantly different side lengths ($L=3~\mathrm{mm}$ \& $L=100~\mathrm{\mu m}$), considering thermomechanical noise $\sigma_{A,TM}$ (dashed lines) and thermal fluctuation noise $\sigma_{A,TF}$ (solid lines). (c) Noise equivalent power (NEP) of SiN membranes of various side lengths $L$ with fixed thickness of 90 nm, considering thermomechanical (TM) and thermal fluctuation (TF) noise. Three sets of traces represent NEP of resonator at different driven vibration amplitudes. (d) Specific detectivity $D^*=\sqrt{A_{det}}/\mathrm{NEP}$ of SiN membrane resonators considering both thermomechanical and thermal fluctuation noises at different vibration amplitudes. All calculations consider a total emissivity of $\varepsilon=0.1$ and absorption $\gamma=0.4$ at the detection frequency ($ \thicksim 2$ THz in this work).}
\end{figure*}

\section{Noise optimization}
For optimizing $\mathrm{NEP}$ of membrane resonator-based radiation sensors, Eq.~(\ref{eq:1}) outlines that we can either seek to minimize $S_y$ or maximize $R$. We first investigate the contribution of $R$, assuming $H_{th,eff}=1$ (i.e., sampling time $\tau \gg \tau_{th}$). As shown in Fig. 1(a), we find that $R$ is optimized for smaller membranes, but only with modest gains at sub-mm dimensions. In Eq.~(\ref{eq:4}), we evaluate $G$ using a closed-form heat transfer model \cite{ref5} that depends on the resonator dimensions (side length $L$, thickness $t$), on the material thermal conductivity ($k=2.7~\mathrm{W/m\cdot K}$) and on the membrane hemispherical total emissivity $\varepsilon$ of approximately 0.1 for a plain 90-nm-thick SiN \cite{ref5}. In turn, for the first few mechanical modes, $\alpha$ is well approximated by \cite{ref5}:  
\begin{equation}
\alpha \cong \frac{E \alpha_T}{2\sigma(1-\nu)},
\label{eq:6}
\end{equation}
where $E=300~\mathrm{GPa}$ is Young’s modulus, $\alpha_T = 2.2\times10^{-6}~\mathrm{K^{-1}}$ is the membrane material thermal expansion coefficient, $\sigma=100~\mathrm{MPa}$ is the built-in tensile stress, and $\nu=0.28$ is the Poisson ratio. For higher order modes, Eq.~(\ref{eq:6}) generally yields an error of less than 20\%, which is detailed in previous works \cite{ref5,ref20}. Thus, for a given set of material constants, the dependency of $R$ on the dimensions of the membrane resonators arises solely from the dimensional dependency of $G$. As shown in Fig. 1(a), minimizing $L$ increases the membrane responsivity by reducing thermal radiative heat transfer with the environment. However, such improvement gradually plateaus for sub-mm values of $L$, when $G$ becomes strongly dependent on solid-state conduction and weakly dependent on $L$. Extended discussion on the contribution of radiation and conduction heat transfer in membranes can be found in previous works \cite{ref5,ref15}. 

In contrast to $R$ varying weakly with membrane dimensions at $L < 1~\mathrm{mm}$, we find that frequency noises can be minimized significantly using membrane resonators with large dimensions (i.e., $L > 1~\mathrm{mm}$). We compute theoretical Allan deviation $\sigma_A$ (see Eq. (S3) in Supplementary Section S2) from expected fractional frequency noise spectral density $S_{y,TF}$ and $S_{y,TM}$. In Fig. 1(b) we compute, as an example, the theoretical $\sigma_A$ of 90-nm-thick SiN membrane resonators of significantly different sizes (i.e., $L = 3$ $\mathrm{mm}$, and $L = 100~\mathrm{\mu m}$), considering actuation of mode (1,1) at an arbitrary low drive amplitude (i.e., 10\% of the critical drive given by $A_{crit}=0.56~L\sqrt{\sigma/QE}$ \cite{ref20}). We scale $m_{eff}$, $f_r$, $Q$, $A_{crit}$ and $G$ with the dimension $L$ using the relations given in Supplementary Section S3, from which we consider $Q\approx3\times10^5$ and $1\times10^6$ for membranes of $L = 100~\mathrm{\mu m}$ and 3 $\mathrm{mm}$. For the large membrane, we obtain $\tau_{mech}\gg\tau_{th}$ such that thermomechanical noise ($\sigma_{A,TM}$) is significantly filtered (i.e., attenuated) when sampling at the resonator thermal time constant $\tau_{th}$, i.e., when sampling as fast as the membrane can thermally respond ($\tau_{th}=100$ ms). This is not the case for the smaller membrane, in which thermomechanical noise is a dominant noise source that adds to thermal fluctuations. As a result, at $\tau_{th}=100$ ms, the total noise in the large membrane ($\sigma_A\approx\sigma_{A,TF}\approx1\times10^{-9}$) is two orders of magnitude lower than in the small membrane ($\sigma_A\approx\sigma_{A,TM}\approx1\times10^{-7}$).

Finally, we strike a balance between optimizing responsivity $R$ and frequency stability, from which we find that the optimal resonator dimensions are on the order of 1 $\mathrm{mm}$. Fig. 1(c) presents a more comprehensive view on resonators $\mathrm{NEP}$ over a range of commonly used sizes (i.e., $L = 50~\mathrm{\mu m}$ and 10 $\mathrm{mm}$). Here, we calculate $\mathrm{NEP}$ directly using Eq.~(\ref{eq:1}) considering $S_y=S_{y,TF}+S_{y,TM}$. We note that $\mathrm{NEP}$ deteriorates at both extreme large and small $L$. As $L$ becomes smaller than 1 $\mathrm{mm}$, the $\mathrm{NEP}$ is affected by excessive levels of thermomechanical noise, unless driven at an amplitude close to $A_{crit}$, which is often challenging in practice \cite{ref33,ref34,ref35,ref36}. Conversely, as $L$ gets exceedingly large (i.e., $L>3~\mathrm{mm}$), $\mathrm{NEP}$ is harmed by diminishing $R$ (see Fig. 1a). Consequently, as shown in Fig. 1(c), $\mathrm{NEP}$ is inherently optimal within the range $1~\mathrm{mm}<L<3~\mathrm{mm}$. Interestingly, a similar optimal range was recently observed experimentally in \cite{ref25} by systematic testing of membranes of several dimensions. We then normalize the $\mathrm{NEP}$ values presented in Fig.~1(c) by $\sqrt{A_{det}}=L$ to obtain $D^*$ in Fig.~1(d) and compare it to our photon fluctuation-limited $D^*_\mathrm{TF,~photon}\approx1.6\times10^{10}~\mathrm{cm\cdot\sqrt{Hz}/W}$. We note that when $L < 1~\mathrm{mm}$, $D^*$ degrades quickly to below $D^*_\mathrm{TF,~photon}$ due to excessive thermomechanical noise and plateauing thermal responsivity as membrane size decreases. In contrast, at larger $L$, $D^*$ gradually approaches $D^*_\mathrm{TF,~photon}$ due to thermal conductance being dominated by radiation ($G\approx G_{rad}$).

\section{Experimental methods}

Building on this analysis we opt for a relatively large $3.2\times3.2$ $\mathrm{mm}$ square SiN membrane resonator as the sensing platform, which we functionalize with a 70-nm-thick titanium metasurface to enable THz absorption. The THz absorption spectrum of the metasurface is designed using Ansys Lumerical finite-difference-time-domain (FDTD) simulation software, from which we expect a peak $\gamma_{peak}=0.4$ at $ \sim2$ THz, as shown in Fig. 2(a). We deposit this metasurface with a 1-$\mathrm{mm}$ diameter $D$ at the center of the membrane (see Fig. 2a) to ensure a sufficiently large peripheral area of plain SiN for optomechanical interrogation. This metal-free region prevents the interrogation laser source (i.e., a 1564 nm distributed feedback laser) from impinging and heating the titanium metasurfaces which could degrade resonators frequency stability due to laser fluctuation \cite{ref14,ref16,kanellopulos2024stress}. 

By covering only part of the membrane resonator, we need to adjust our responsivity, $\mathrm{NEP}$ and $D^*$ calculation from those of a uniform membrane in Section II. We therefore solve a combined modes heat equation (i.e., coupled radiation and conduction) of the SiN membrane resonator numerically in MATLAB by defining a heating zone at the geometric center of the membrane, which accounts for the effective localized heating area (i.e., effective diameter of the titanium metasurface $D$). Fig. 2(b) exhibits the variation in $R$ when a $D=1$ $\mathrm{mm}$ localized titanium metasurface incorporated at the geometric center of a SiN membrane resonator at different sizes $L$. We find that our approach (i.e., $D=L/3=1$ $\mathrm{mm}$) sacrifices $\sim 40\%$ of $R$, compared with a uniformly heated (i.e., $D=L=3$ $\mathrm{mm}$) SiN membrane resonator. Using this adjusted $R$ and the thermal fluctuation noise $S_{y,TF}$ for a full membrane (See Eq.~\ref{eq:3}), we predict theoretical minimum $\mathrm{NEP\approx13~pW/\sqrt{Hz}}$ for our sensor geometry in Fig.~2(a). In addition, using detector area $A_{det}\approx0.79~\mathrm{mm^2}$, we predict $D^*\approx7\times10^9~\mathrm{cm\cdot\sqrt{Hz}/W}$ as the theoretical maximum detectivity for this particular geometry.

\begin{figure}[!htb]
\includegraphics[scale=0.95]{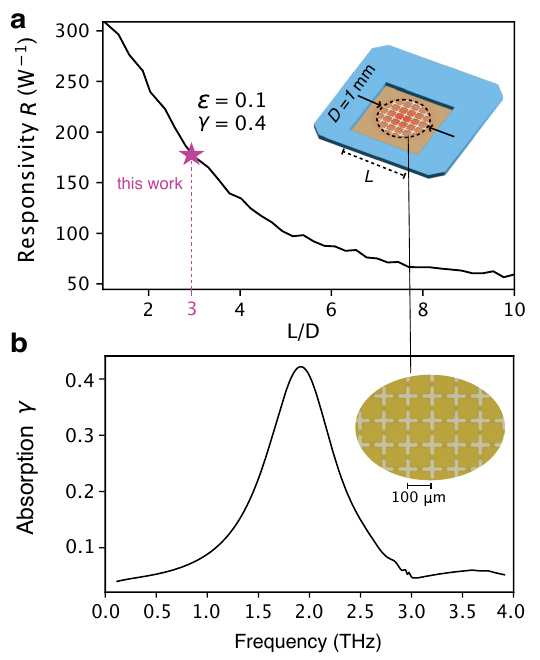}
\caption{\label{fig:2}(a) Numerically computed thermal responsivity $R$ of a $D=1$ $\mathrm{mm}$ diameter localized THz metasurface absorber incorporated on top of SiN membrane resonators of various side lengths $L$. The purple star indicates the responsivity of the fabricated device. (b) Numerically computed absorption spectrum of the fabricated metasurface in the THz frequency range. Inset: microscope photograph of the fabricated metasurface.}
\end{figure}

The plain SiN membrane resonator is fabricated in-house using a 90-nm-thick low-pressure chemical vapor deposition (LPCVD) low-stress SiN-on-silicon wafer. The titanium is deposited onto the surface of the SiN membrane resonator via electron beam evaporation through a custom-made shadow mask to form the metasurface. Detailed fabrication processes are described in previous work \cite{ref26}.

\begin{figure*}
\includegraphics[scale=0.58]{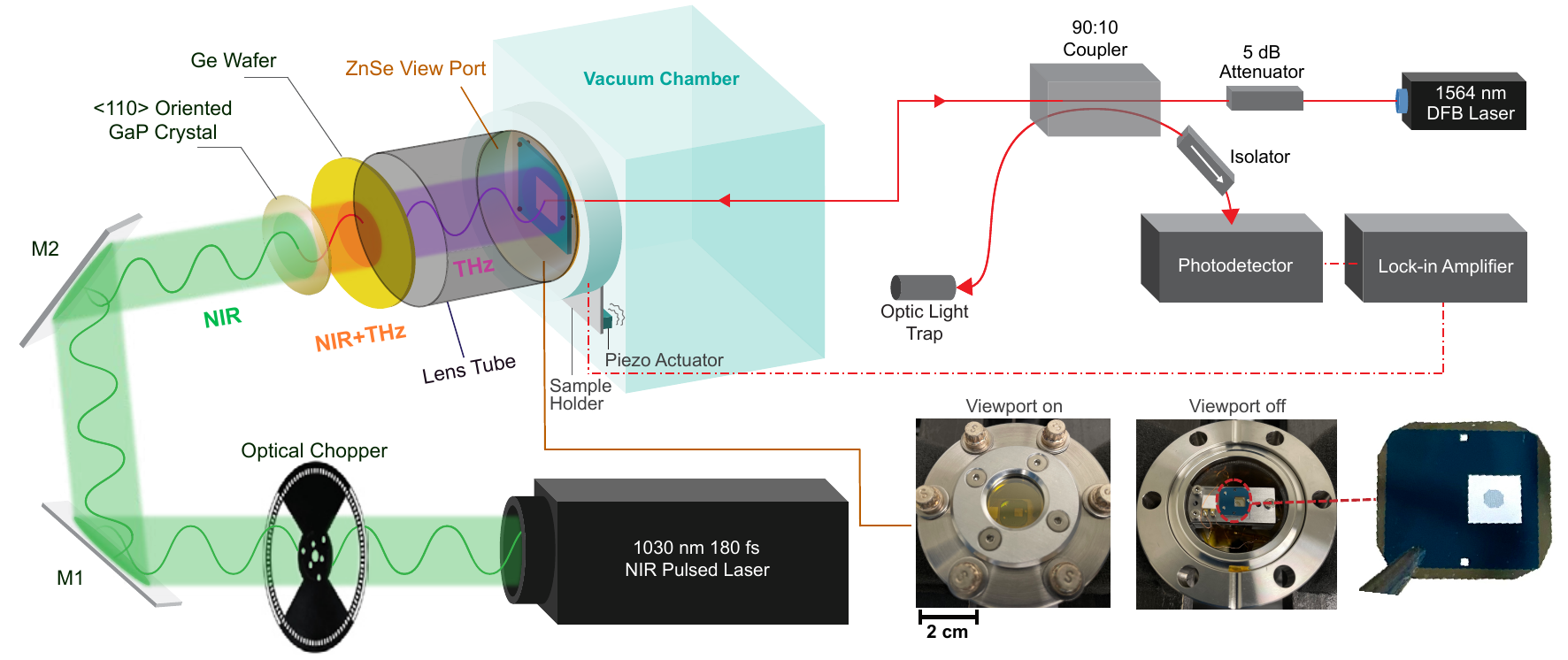}
\caption{\label{fig:3}Schematics of the overall experimental setup which includes a laser interferometer outside of the vacuum chamber for resonator readout, SiN membrane resonator mounted inside the chamber, and terahertz nonlinear generation process using near infrared (NIR) pulsed laser incident on a gallium phosphite GaP crystal.}
\end{figure*}

Once fabricated, the resonator is placed in a custom-made portable high vacuum ($\sim10^{-6}$ hPa) chamber \cite{ref27} to minimize convective heat transfer and damping by air. The resonator is mounted on a steel plate by three pairs of disc magnets and mechanically excited via a piezo actuator. The membrane is aligned with the center of a Zinc Selenide (ZnSe) view port for easy optical alignment (See Fig. 3). A single-mode optical fiber tip is pointed at the back side of the resonator for optical interrogation of its mechanical vibration. We probe vibration signal of the resonator using a custom-assembled laser interferometer \cite{ref28} located outside the vacuum chamber and consisting of a 1564 nm $\mathrm{Orion^{TM}}$ distributed feedback (DFB) laser, a 5 dB optical attenuator, a 90:10 coupler and a Thorlabs PDA20CS2 photodetector. The combined use of the optical attenuator and coupler reduces the laser power to 11.7 $\mathrm{\mu W}$ before reaching the SiN membrane resonator. This largely attenuated laser power produces sufficient signal for detection, while preventing any noticeable laser fluctuation from degrading resonator frequency stability \cite{ref14,ref16,kanellopulos2024stress}.

Choosing an ideal mechanical mode for thermal sensing requires simultaneously optimizing three parameters: $f_r$, $Q$ and $A_{rss}$, to suppress thermomechanical noise $S_{y,TM}$ as Eq.~(2) indicated. We therefore first perform a frequency sweep to locate the optimum mechanical mode of our sample. Note that the optimum mode must be determined experimentally since these three parameters are sensitive to variations in sample mounting conditions. During our experiment, we chose a high Q-factor ($Q=870,000$) mechanical eigenmode (i.e., mode order 2,3) at 124 kHz. This eigenmode also exhibits a higher demodulated signal amplitude (i.e., higher $A_{rss}$) compared with other modes under the same actuation signal. Additional information regarding other mechanical modes can be found in Supplementary Section S6. We use a Zurich Instrument Ltd. MFLI lock-in amplifier (LIA) to excite our sample at this mode below the critical amplitude $A_{crit}$ to avoid frequency stability degradation due to nonlinearity. We track its resonance frequency shift upon THz light absorption via a built-in phase-locked loop (PLL) frequency tracking function. We set both demodulation bandwidth (5 kHz), and sampling rate (32,000 Sa/S) to very high values, which we can numerically average to lower effective sampling rates in post processing. We set the PLL bandwidth to 8 Hz, which ensures that the PLL tracking speed is roughly five times faster than the thermal time constant $\tau_{th}\approx100$ ms) of our plain $3.2\times3.2$ mm SiN membrane \cite{ref15,ref16}, such that the true thermal response of the SiN membrane resonator can be recorded without filtering.

We generate collimated THz radiation with spectrum centered around 1.8 THz via optical rectification of a collimated near-infrared (NIR) pulsed laser beam (1 mJ pulse energy, 180 fs pulse duration, 6 kHz pulse repetition rate) in a 2-mm-thick <110>-oriented gallium phosphide (GaP) crystal (see Fig. 3). The generated THz radiation is pulsed with the same repetition rate but is perceived as CW by our sensor of comparatively slow response time ($\tau_{th}\approx200$ ms). Similar THz generation process is detailed in Cui et. al \cite{ref29}. A germanium (Ge) wafer that is transparent to THz radiation is placed in the optical path (see Fig. 3) to block the residual NIR light, ensuring that only the THz light can reach the SiN membrane resonator. Additionally, a 20-cm long, circular hollow lens tube is positioned between the Ge wafer and the ZnSe viewport of the portable vacuum chamber (see Fig. 3), to prevent any possible external stray light from reaching the sample. This relatively long (20 cm) propagation length also geometrically attenuates, via divergence, parasitic thermal radiation generated in the Ge wafer due to NIR absorption, while the coherent THz beam remains collimated to an approximately constant beam diameter (6 mm).
\begin{figure}[!htb]
\includegraphics[scale=0.225]{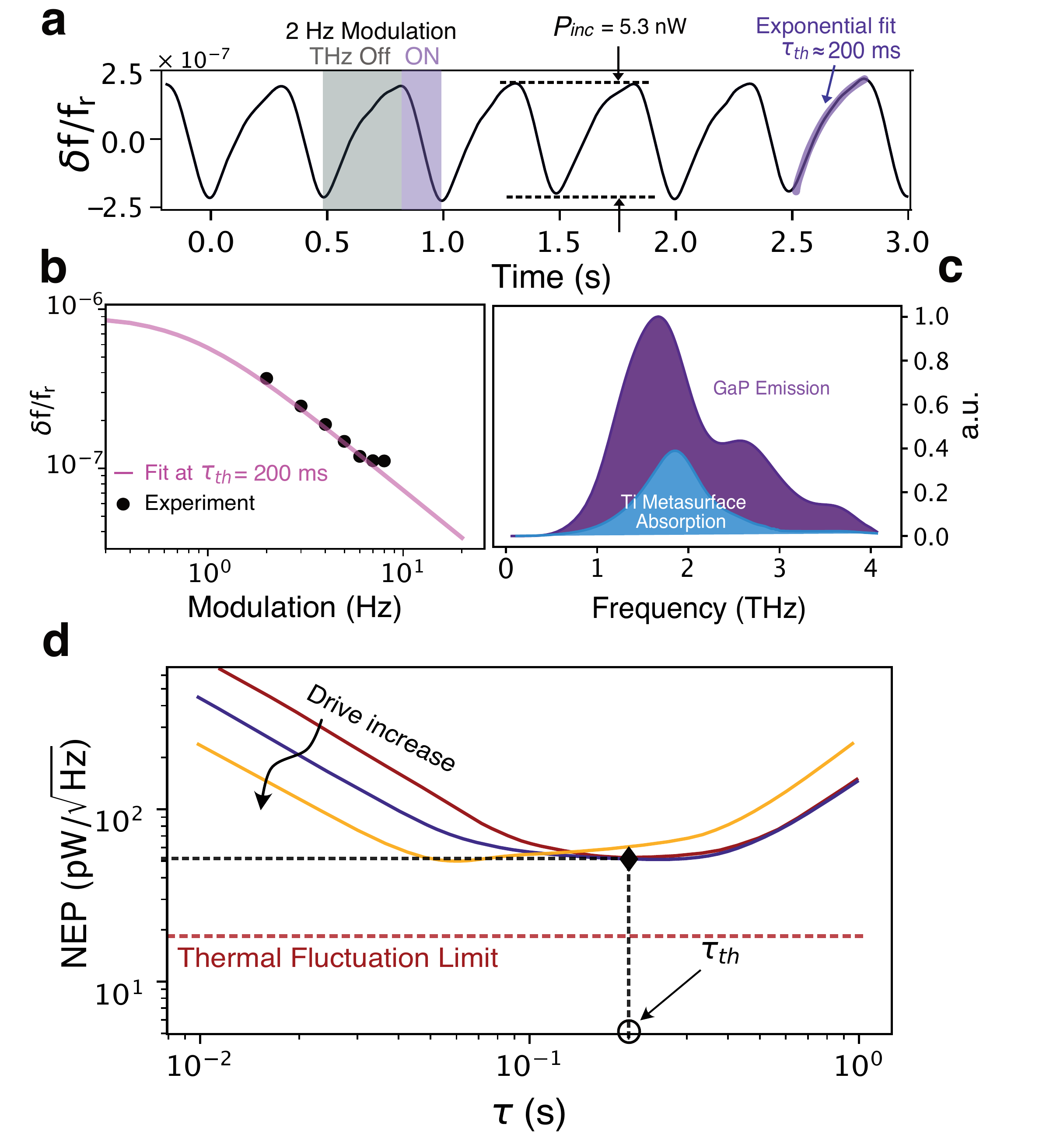}
\caption{\label{fig:2}(a) Experimental fractional frequency shift $\delta f/f_r$ time trace upon terahertz absorption at 2 Hz optical modulation and filtered at phase-lock loop (PLL) bandwidth of 8 Hz. The exponential fit shown in purple line yields the thermal time constant $\tau_{th}$. (b) $\delta f/f_r$ amplitude at various optical modulation frequencies. (c) Normalized experimental GaP emission power spectrum matched with simulated absorption power spectrum of the Ti metasurface. (d) Experimental noise equivalent power ($\mathrm{NEP_{eff}}$) of our device at different drive amplitudes, calculated from experimental Allan deviation $\sigma_A$ and thermal responsivity $R$.}
\end{figure}

\section{Results~\&~Discussion}
We first validate the heat transfer model in our resonators by recording the fractional mechanical frequency shifts $\delta f/f_r$ in our sample when exposed to a 6-mm-diameter THz beam, modulated at 2 Hz via an optical chopper. This is shown in Fig. 4(a), in which the effective sampling rate is set to match the PLL bandwidth (8 Hz). From this figure, we obtain an experimental thermal response time $\tau_{th}\approx200$ ms that is roughly two times larger than the expected $\tau_{th}$ of a plain SiN membrane \cite{ref15,ref16}. This is in close agreement with finite element simulations (see Supplementary Section S4), that predicts reduced response speed due to THz absorption occurring in a localized region, and the additional thermal mass of the titanium metastructures. We then repeat the same experiment at different optical modulation frequencies (from 2 Hz to 8 Hz), from which we obtain the expected frequency roll-off a 1-pole low pass filter of thermal time constant $\tau_{th}$ (see Fig. 4b).

We then measure the effective optical absorption ($\gamma_{eff}\approx27\%$) of our metasurface for the specific THz source used in for our experiment, which allow us to extract the sensor responsivity. We perform electro-optic sampling (EOS) \cite{ref29} to measure the THz emission power spectrum generated by non-linear conversion in the GaP crystal (see Fig. 3c, purple curve). We then compare this emission spectrum to the absorption spectrum of our metasurface (see Fig. 2b). This comparison is shown in Fig.~4(c), from which we infer that our metasurfaces absorbs $\gamma_{eff}\approx27\%$ of the incident THz light for the source used in the present experiments (i.e., for a source frequency spanning from 0.5 THz to 4 THz, see Fig.~4c). This is different than the peak absorption of $\gamma_{eff}\approx40\%$ at our designed frequency (2 THz in Fig. 2b). We therefore adjust our predicted responsivity by 33\% from the value predicted in Fig. 2(a), i.e., we use $R\approx120~\mathrm{W^{-1}}$ in the following. Therefore, the theoretical performance limit can be adjusted by the same 33\% as $\mathrm{NEP_{\mathrm{eff}}\approx19~pW/\sqrt{Hz}}$ and $D^*_{\mathrm{eff}}\approx4.7\times10^9~\mathrm{cm\cdot\sqrt{Hz}/W}$.

Using this responsivity, we can relate the measured fractional frequency shift $\delta f/f_r$ in Fig. 4(a), to the THz power incident on our metasurface using 
\begin{equation}
P_{inc} = \frac{\delta f/f_r}{R}.
\label{eq:9}
\end{equation}
This yields an incident power of 5.3 nW as indicated in Fig. 4(a). Likewise, we can normalize this incident power by the area of our metasurface ($\mathrm{\pi} D^2/4=0.79~\mathrm{mm^2}$) to estimate the average intensity of the THz light to 6.7 $\mathrm{nW/mm^2}$ at the metasurface location. We also estimate that $\approx40\%$ of the generated THz light transmits through the ZnSe viewport \cite{ref30}, such that the incident intensity prior to entering our vacuum chamber is $\approx16.7~\mathrm{nW/mm^2}$. Interestingly, this value is close agreement with a comparison measurement taken using a Tydex GC-1D Golay cell, which yields a $\approx16~\mathrm{nW/mm^2}$ measured intensity. 

Using this confirmed responsivity ($R=120~\mathrm{W^{-1}}$), we can estimate our detector $\mathrm{NEP}$ by measuring the Allan deviation noise trace ($\sigma_A$) in the absence of incident THz radiation, and then using \cite{ref2,ref6,ref8}

\begin{equation}
\mathrm{NEP}=\frac{\sigma_A(\tau)\cdot\sqrt{\tau}}{R}. 
\label{eq:10}
\end{equation}

From this, we obtain Fig. 4(b), which indicates a minimum $\mathrm{NEP_{eff}}\approx51~\mathrm{pW/\sqrt{Hz}}$ at a sampling time $\tau_{th}$, for our specific broadband terahertz source spanning 0.5--4 THz. Correspondingly, we obtain $\mathrm{NEP_{peak}}\approx36 ~\mathrm{pW/\sqrt{Hz}}$ and $D^*=3.4\times10^9~\mathrm{cm\cdot\sqrt{Hz}/W}$ at our central metasurface design wavelength (2 THz) where absorption is $\gamma_{peak}\approx0.4$. In addition, we plot theoretical $\mathrm{NEP_{eff}}\approx19~\mathrm{pW/\sqrt{Hz}}$ in Fig. 4(d), indicating the ideal performance of detector when being limited by thermal fluctuation $S_{y,th}$. Our experimental $\mathrm{NEP_{eff}}$ is therefore only a factor of 3 from this limit. Reducing this factor will likely be readily possible in future experiments, as we previously achieved frequency noise limited by thermal fluctuations in \cite{ref16} using a different experimental apparatus. In the present case, mechanical limitations of our portable vacuum chamber most likely caused non-idealities in our displacement readout interferometer. In Fig. 4(d), at $\tau\ll\tau_{th}$, the uptick of our experimental $\mathrm{NEP_{eff}}$ is caused by noise in our optical readout, combined with $R$ approaching zero for decreasing $\tau$ in the denominator of Eq.~(1). At $\tau\gg\tau_{th}$, $\mathrm{NEP_{eff}}$ is negatively affected by systematic drift.

We finally test our detector with attenuated incident THz optical power to test its linearity. We employ a metasurface thin-film polarizer in front of the ZnSe viewport of the vacuum chamber for varying the intensity of the linearly polarized THz light. When rotating the polarizer away from the maximum transmission orientation by an angle $\theta$, we expect the transmitted THz power to vary by a factor $\mathrm{sin}(\theta)^2$. This is recovered exactly in Fig. 5(a), for both our sensor and the control Golay cell. This exact correspondence with the Golay cell and the expected $\mathrm{sin}(\theta)^2$ signal attenuation confirms the linearity of our sensor, which is better illustrated by plotting the same measured SiN signal against the incident THz power in Fig. 5(b). We also note that the polarization sensitivity observed in Fig. 5 rules out the possibility that we are detecting thermal light emitted from the NIR-absorbing Germanium wafer, which would be unpolarized.

\begin{figure}[!htb]
\includegraphics[scale=0.9]{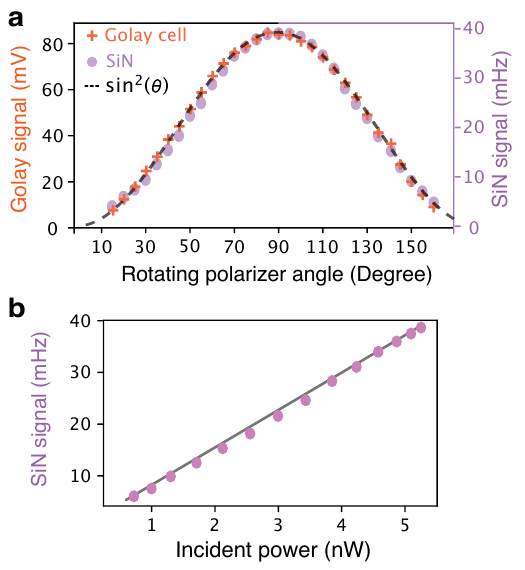}
\caption{\label{fig:5}(a) Comparison experiment between a commercial Golay cell signal and SiN membrane resonator frequency shift signal $\delta f$. The polarizer angle attenuates the incident optical power by a factor $\mathrm{sin}(\theta)^2$. (b) SiN membrane resonator frequency shift signal $\delta f$ as a function of incident power retrieved from (a).}
\end{figure}

The same attenuation experiment confirms the performance of our sensor at low optical power. Fig. 6 presents the same response as in Fig. 4(a), but at different attenuation power (i.e., polarizer angle $\theta$). We confirm that our sensor can clearly detect optical power of 0.7 nW, which was expected by our measured $\mathrm{NEP_{eff}}\approx51~\mathrm{pW/\sqrt{Hz}}$ and data sampling rate of 8 Hz, from which the expected detection limit is 0.14 nW.

\begin{figure}[!htb]
\includegraphics[scale=1]{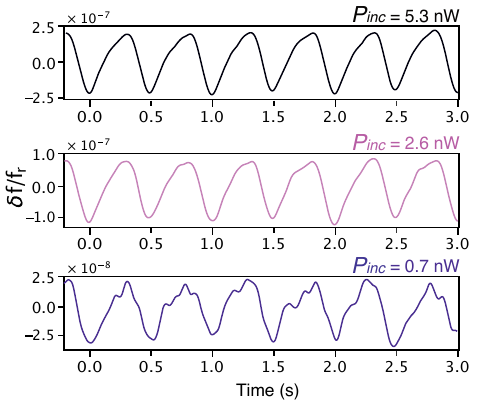}
\caption{\label{fig:5}Fractional frequency shift $\delta f/f_r$ time trace of the SiN membrane resonator at various levels of attenuated incoming THz power, confirming performances at optical power as low as 0.7 nW.}
\end{figure}

\section{Conclusion}
We show that, when noise sources are well estimated and designed-for, high performance radiation sensing at long optical wavelengths is possible using ubiquitous square SiN membrane resonators. Using localized terahertz metasurfaces absorbers, a peak detectivity of $3.4\times10^9~\mathrm{cm\cdot\sqrt{Hz}/W}$ at around 2 THz is experimentally demonstrated. Such detectivity has not been previously realized by any existing nanomechanical resonator, nor by any commercial room-temperature on-chip terahertz detectors. More generally, it is on par with recent record-breaking devices at infrared wavelengths \cite{ref31,ref32}. With improvement of our optical readout method already achieved in \cite{ref16}, the performance of our device can feasibly be improved by another factor of 3. Achieving such high level of detectivity while operating at room temperature makes this nanomechanical resonator THz detector particularly interesting for applications that require high-performance single-point detection, such as far-infrared spectroscopy, where response time is often less critical than high detectivity and room-temperature operation.

\section*{Supplementary Material}
See supplementary S1 for fractional frequency noise spectral density in angular frequency $\mathrm{rad/s ^{-1}}$, supplementary S2 for transformation from fractional frequency noise spectral density $S_y(f)$ to Allan deviation $\sigma_A$, S3 for scaling noise equivalent power NEP parameters with SiN membrane resonator sizes, S4 for defining the effective heat transfer area of SiN membrane resonator during terahertz absorption, S5 for estimating thermal responsivity $R$ under localized heating, S6 for additional mechanical modes information. 

\begin{acknowledgments}
The author would like to acknowledge Minji Choi for creating some of the graphical illustrations included in the figures of this article.
\end{acknowledgments}

\section*{references}
\bibliography{aipsamp}

\end{document}